\documentclass[twocolumn]{aastex631}
\usepackage{amsmath,amssymb, amsthm,amstext}
\usepackage{natbib}
\usepackage{graphicx}
\usepackage{color}
\usepackage{array, enumerate}
\usepackage{bm}
\usepackage{multirow}
\usepackage{braket}
\usepackage{txfonts}
\usepackage{physics}
\usepackage[normalem]{ulem}

\newcommand{\blackdiamond}{\rotatebox[origin=c]{45}{$\vcenter{\hbox{$\blacksquare$}}$}}

\def\be{\begin{equation}}
\def\ee{\end{equation}}

\begin{document}

\title{A Formation Crisis of Repeating Partial Tidal Disruption Events}

\author[0000-0001-9608-009X]{Zhen Pan}
\email{zhpan@sjtu.edu.cn}
\affiliation{Tsung-Dao Lee Institute, Shanghai Jiao-Tong University, Shanghai, 520 Shengrong Road, 201210, People’s Republic of China}
\affiliation{School of Physics \& Astronomy, Shanghai Jiao-Tong University, Shanghai, 800 Dongchuan Road, 200240, People’s Republic of China}

\author[0000-0002-1934-6250]{Dong Lai}
\email{donglai@sjtu.edu.cn}
\affiliation{Tsung-Dao Lee Institute, Shanghai Jiao-Tong University, Shanghai, 520 Shengrong Road, 201210, People’s Republic of China}
\affiliation{School of Physics \& Astronomy, Shanghai Jiao-Tong University, Shanghai, 800 Dongchuan Road, 200240, People’s Republic of China}
\affiliation{Department of Astronomy, Center for Astrophysics and Planetary Science, Cornell University, Ithaca, NY 14853, USA}

\begin{abstract}
A number of candidate repeating partial tidal disruption events (rpTDEs) have been reported in recent years.
If these events are confirmed, the high fraction of observed rpTDEs among all tidal disruption events (TDEs) is in tension with prediction of the single-star loss-cone channel as noticed in previous studies. We further point out an inequality $M_\bullet \lesssim 4\times 10^6 M_\odot (T_{\rm obt}/10\ {\rm yr})^{4/9}$ that must be satisfied
for rpTDEs of solar type stars in the loss-cone channel, where $M_\bullet$ is the central supermassive black hole (SMBH) mass and $T_{\rm obt}$ is the orbital period of the star. However the majority of reported rpTDE candidates potentially violate this inequality, indicating an alternative formation channel. For the commonly invoked Hills mechanism to explain the rpTDE observations, the majority of tidally disrupted stellar binaries by  SMBHs must be near-contact binaries. 
If the same process operates in the Galactic Center, there
should exist a population of hypervelocity stars (HVSs)
ejected with velocities as high as $3.6\times 10^3 (M_\bullet/10^6 M_\odot)^{1/6}\ {\rm km\ s}^{-1}$, which however have not been detected.
A complete search for HVSs in the Milky Way will be critical for testing this prediction.
\end{abstract}

\section{Introduction} \label{sec:intro}

When a star approaches a supermassive black hole (SMBH) too close, it is tidally disrupted and approximately half of the stellar debris falls back to the SMBH, resulting in accretion and outflow, and producing a bright flare \citep{Hills1975,Ress1988}. In the classical loss cone theory, the rate of tidal disruption events (TDEs) is determined by the relaxation process in which stars are kicked onto highly eccentric orbits with the pericenter distance $r_{\rm p}$ less than the tidal disruption radius $r_{\rm t}\equiv (M_\bullet/m_\star)^{1/3} R_\star$  \citep{Alexander2017ARA&A}, where $M_\bullet$
is the mass of the SMBH, $m_\star$ and $R_\star$ are the mass and radius of the star, respectively. The computed TDE rate of $(1-2)\times 10^{-4} \text{ galaxy}^{-1} \text{ yr}^{-1}$ for Milky Way-like galaxies, derived from quasi-empirical models \citep{Wang2004, Vasiliev2013, Stone:2014wxa}, is slightly higher than the observed rate of a few $\times 10^{-5} \text{ galaxy}^{-1} \text{ yr}^{-1}$ inferred from ZTF \citep{Yao:2023rbr}. Recent studies of nearby galaxies with high-resolution nuclear density measurements show that the sample-averaged TDE rate derived from the loss-cone theory agrees well with observations \citep{Hannah:2024dmr}, and the theoretical rate prediction for
Milky-Way-mass galaxies is also consistent with the observed rate estimate \citep{Hannah:2024dmr, Chang:2024hdb}. However, the TDE rate inferred from optical surveys may miss a substantial fraction of events, particularly those obscured by dust, which can be instead unveiled by infrared surveys~\citep{Jiang2021,Masterson2024}.

A star could be partially disrupted if its pericenter distance $r_{\rm p}$ is a few times larger than the tidal disruption radius $r_{\rm t}$.
The partial TDE (pTDE) rate in the loss-cone theory has been investigated in several previous studies, extending from the so-called full loss cone regime \citep{Stone:2020vdg} to the empty loss cone regime \citep{Krolik2020,Chen2021,Zhong2022,Bortolas:2023tlq}.
A subtle point of counting TDEs is that pTDEs involving the same star can occur multiple times, and such multiple disruptions can increase the “observed" TDE rate if they are counted as independent events \citep{Zhong2022,Bortolas:2023tlq}. \cite{Broggi:2024qkj} estimated the number of consecutive partial disruptions that a star can undergo accounting for mass loss and tidal excitation during each pericenter passage; the authors claimed that the pTDE rate can be higher than that of full TDEs by a factor of a few or more.

In the last few years, a number of repeating partial TDE (rpTDE) candidates have been reported in the literature, with the interval between the consecutive flares ranging from a few months to tens of years \citep{Payne:2020tfp,Liu:2022avb,Lin:2024reb,Sun2024,Somalwar:2023sml,Wevers:2022tsr,Veres:2024qcm,Ji:2025brt,Malyali:2023xnl,Sun2024,
Wang:2025oyc}.\footnote{Note that among these candidates only AT 2022dbl is a spectroscopically confirmed, optically selected TDE \citep{Lin:2024reb,Makrygianni2025}. } The observations of rpTDEs provide an important probe of the formation mechanism of TDEs. 
In fact, not all rpTDEs are observable since weak encounters (with large $r_{\rm p}$) may lead to negligible stellar mass loss $\Delta m$ and produce dim flares that are undetectable, and rpTDEs with long intervals between flares (i.e. long stellar period $T_{\rm obt}$)
are easily identified as full TDEs without sufficiently long-term monitoring. In this paper, we focus on “observable” rpTDEs, i.e., those with sufficiently large mass loss $\Delta m$ (and correspondingly large flare luminosity) and short flare interval (small $T_{\rm obt}$).
For definiteness, we adopt the thresholds 
$\Delta m/m_\star \gtrsim 0.1$ (see Section 2) and $T_{\rm obt}$ less than a few years.

This letter is organized as follows. In Section~\ref{sec:loss cone}, we start with a brief review of loss cone theory, calculate the expected rpTDE rate and the expected distribution of rpTDEs in the $T_{\rm obt}-M_\bullet$ space, and point out discrepancies in both the predicted rate and the distribution compared to the rpTDEs reported in the literature. In Section~\ref{sec:Hills}, we compare prediction of the Hills mechanism (binary breakup) with the rpTDE observations and find they are compatible only if the stellar binaries in nuclear stellar clusters are near-contact binaries. 
However, hypervelocity stars  founded in the Milky Way 
are from detached stellar binaries.
We conclude in Section~\ref{sec:conclusion}.
We use geometrical units with $G=c=1$ throughout this paper.

\section{Tidal Disruptions in Loss cones}\label{sec:loss cone}

\subsection{Basics}\label{sec:basics}

We consider a SMBH with mass $M_\bullet$ in the center of nuclear stellar cluster consisting of stars with mass $m_\star$.
After several times of relaxation timescale \citep{Spitzer1987}
\be 
\begin{aligned}
    t_{\rm rlx} &= \frac{0.339}{\ln\Lambda}\frac{\sigma^3(r)}{m_\star^2n_\star(r)} \\
    & = 2.8\ {\rm Gyr}\ \left(\frac{\ln\Lambda}{10}\right)^{-1} \left(\frac{\sigma}{60\ {\rm km\ s}^{-1}}\right)^3\left(\frac{n_\star}{10^5\ {\rm pc}^{-3}}\right)^{-1}\ ,
\end{aligned}
\ee
nuclear stellar clusters that have relaxed to a steady state,
where $\sigma(r)$ is the local velocity dispersion and is approximately $\sqrt{M_\bullet/r}$ within the radius of influence ($r_{\rm h}$) of the SMBH, and $\ln\Lambda\approx10$ is the Coulomb logarithm.  According to this estimate, Milky Way like galaxies have marginally enough time to establish a steady state. Nuclear stellar clusters that have relaxed to a steady state follow the Bahcall-Wolf density profile $n_\star(r)\propto r^{-7/4}$ \citep{Bahcall1976}.  We will first investigate different types of star disruptions in a Bahcall-Wolf stellar cluster, then discuss the influence of 
deviations in the density profile.

Assuming the SMBH mass and the nuclear stellar cluster velocity dispersion
$\sigma_\star=\sigma(r_{\rm h})$ satisfy the relation
$M_\bullet\approx 10^8M_\odot\,(\sigma_\star/200\,{\rm km\,s}^{-1})^4$
\citep{Tremaine2002,Gultekin2009}, 
the radius of influence of the SMBH is given by
\be 
r_{\rm h}:=\frac{GM_\bullet}{\sigma^2_\star}\approx  M_{\bullet,6}^{0.5}\ {\rm pc}=2.1\times 10^7M_\bullet\,M_{\bullet,6}^{-0.5},
\ee
where $M_{\bullet,6}:=M_\bullet/10^6 M_\odot$.
With this relation and the total number of stars within $r_{\rm h}$, $N_\star = M_\bullet/m_\star$, one can derive the normalization factor in the stellar number density $n_\star(r)$. 

\begin{figure}
\centering
\includegraphics[width=0.5\textwidth]{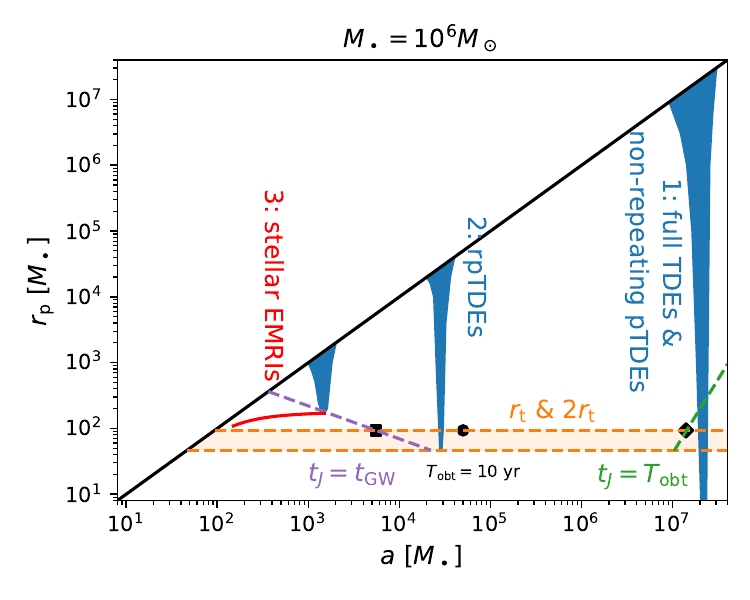}
\caption{Schematic plot showing the formation of full TDEs, non-repeating pTDEs, rpTDEs and stellar EMRIs in a nuclear stellar cluster around a SMBH with $M_\bullet=10^6 M_\odot$,
where $r_{\rm p}$ and $a$ are the pericenter distance and semi-major axis of the stellar orbit, respectively. The line $t_J=T_{\rm obt}$ marks the boundary between the full and empty loss cone regimes. The region on the right-hand side of the black diamond ($\blackdiamond$) dot, $a > a_{\rm full}$ (see Eq.~\ref{eq:full_empty})
is where full TDEs and non-repeating pTDEs are expected to occur.
The line $t_J=t_{\rm GW}$ marks the boundary between GW emission and 2-body scattering dominated regimes. 
The region between the black square ($\blacksquare$) and the black bullet ($\medbullet$) dots, $ a_{\rm GW}(r_{\rm p}=2r_{\rm t}) < a < a(T_{\rm obt}=10\ {\rm yr})$
(see Eq.~\ref{eq:a_JGW}), is where observable rpTDEs (with period less than 10 years) are expected to occur.}
\label{fig:phase_diagram}
\end{figure}

A star in the nuclear cluster is randomly perturbed by gravitational pulls of other stars, and thus diffuses in the orbital energy and angular momentum space. In general, the relaxation timescale in angular momentum is shorter than that in energy ($t_{\rm rlx}$) \citep{Cohn1978,Hopman:2005vr}, i.e.
\be
t_J\simeq (1-e^2)t_{\rm rlx}\ ,
\ee 
where $e$ is the orbital eccentricity. The ratio of $t_J$ and the orbital period $T_{\rm obt}$ is 
\be
\frac{t_J}{T_{\rm obt}}\approx 0.1\left(\frac{r_{\rm p}}{r_{\rm h}} \right)\left( \frac{a}{r_{\rm h}}\right)^{-9/4} \left(\frac{M_\bullet}{m_\star}\right) \ ,
\label{eq:tj}\ee
where $a\sim r$ is the orbital semi-major axis, and we have assumed $r_{\rm p}\ll a$. As a result of two-body scatterings, a star in an eccentric orbit gains and loses its orbital angular momentum randomly while keeping its orbital energy nearly constant 
(see Fig.~\ref{fig:phase_diagram} for illustration; see also
\citet{Sari:2019hot, Linial:2022bjg}). If the star is scattered onto an extremely eccentric orbit with its pericenter distance $r_{\rm p}$ close to the tidal disruption radius 
\be
r_{\rm t} = \left(\frac{M_\bullet}{m_\star}\right)^{1/3} R_\star \approx 46 M_\bullet\ M_{\bullet,6}^{-2/3} m_{\star,\odot}^{-1/3}R_{\star,\odot} \ , 
\ee 
the star is expected to be tidally disrupted by the SMBH, and $r_{\rm p}=r_{\rm t}$ is usually defined as  the loss cone boundary of TDEs. 
As we will show later in Section~\ref{subsec:boundary}, the loss cone boundary actually varies mildly depending on different types of star disruptions.

There are two ways that stars approach the SMBH.
In the full loss cone regime, $t_J < T_{\rm obt}$, the orbital angular momentum of the star is altered by order of unity within one orbital period, i.e., the loss cone is fully occupied by stars that are scattered into or out of the loss cone randomly. These are the sources of full TDEs and non-repeating pTDEs (see label “1: full TDEs $\&$ non-repeating pTDEs” in Fig.~\ref{fig:phase_diagram}). 
In the empty loss cone regime, $t_J > T_{\rm obt}$, the orbital angular momentum of the star is slightly perturbed within one orbital period, 
a star with $1\lesssim r_{\rm p}/r_{\rm t}\lesssim 2$ stays in the partial TDE regime for more than one orbital periods, therefore contributes to rpTDEs (see label “2: rpTDEs” in Fig.~\ref{fig:phase_diagram}). 
Using Eq.~(\ref{eq:tj}), we find that the boundary between full and empty loss cone regimes is given by 
\be\label{eq:full_empty}
\frac{a_{\rm full}(r_{\rm p})}{r_{\rm h}} \approx 0.5\,\beta^{-4/9} M_{\bullet,6}^{10/27} m_{\star,\odot}^{-16/27} R_{\star,\odot}^{4/9}\ ,
\ee 
where we have introduced the penetration factor $\beta:= r_{\rm t}/r_{\rm p}$. Thus, stars with $a\gtrsim a_{\rm full}$ ($a\lesssim a_{\rm full}$) approach the SMBH in the full (empty) loss cone regime.

For a star sufficiently close to the SMBH such that the orbital decay timescale due to gravitational wave
(GW) emission, $t_{\rm GW}$, becomes less than $t_J$,
its orbit circularizes gradually and becomes a stellar extreme mass ratio inspiral (EMRI; see label “3: stellar EMRIs” in Fig.~\ref{fig:phase_diagram}). For $r_{\rm p}\ll a$, we have
\be
{t_{\rm GW}\over T_{\rm obt}}\approx {1\over 31}\left({r_{\rm p}\over M_\bullet}\right)^{5/2}{r_{\rm p}\over a}\left({M_\bullet\over m_\star}\right),
\ee
and (cf. \cite{Linial:2022bjg})
\be\label{eq:tJ_tGW}
\frac{t_J}{t_{\rm GW}}\approx 3 \left( \frac{r_{\rm p}}{M_\bullet}\right)^{-5/2} \left(\frac{a}{r_{\rm h}} \right)^{-5/4}\ .
\ee 
Setting $t_J=t_{\rm GW}$ defines a critical “GW radius":
\be \label{eq:a_JGW}
\frac{a_{\rm GW}(r_{\rm p})}{r_{\rm h}} \approx 1.1\times10^{-3} 
\beta^{2} M_{\bullet,6}^{4/3}\,m_{\star,\odot}^{2/3}\,
R_{\star,\odot}^{-2}.
\ee 
Note that the critical GW radius $a_{\rm GW}$ is insensitive to the assumed stellar density profile $n_\star(r)$ due to the sharp dependence of GW emission timescale on radius $t_{\rm GW} \propto r^4$. 
After a GW circularization stage, a stellar EMRI gradually loses its mass as its pericenter crosses the critical radius $\approx 2r_{\rm t}$ \citep{Dai2013,Lu2023,Linial2017,Linial2023b,Linial2024}. The timescale of mass loss and subsequent accretion into the SMBH of each pericenter passage is comparable to or even longer than the stellar orbital period. As a result, the light curve of this process may not be as dramatic and characteristic as those of TDEs. Although there is no hard boundary separating stellar EMRIs from rpTDEs as shown in Fig.~\ref{fig:phase_diagram} (e.g., if a star enters the GW-dominated regime with its pericenter distance $r_{\rm p}$ slight larger than $2r_{\rm t}$, it will be repeatedly disrupted before the orbit gets substantially circularized), the boundary is actually very sharp since, as we show below, the observed rpTDEs are from highly eccentric stars with $1-e\ll 1$, and these are rare for stellar EMRIs \citep{Linial2023b}.

\subsection{Rates of star disruptions}

At every radius $r$, stars on quasi-circular orbits are scattered onto  highly eccentric orbits and get captured by the SMBH in the form of (partial) TDEs or stellar EMRIs on the relaxation timescale $t_{\rm rlx}(r)$. Consequently, the capture rate in the empty loss cone regime ($a\lesssim a_{\rm full}$) is \citep{Hopman:2005vr}
\be
R_\star(r < a) \approx\frac{N_\star(r < a)}{t_{\rm rlx}(a)} \propto a \ ,
\ee 
where $N_\star(r<a)=4\pi\int^a_0 n_\star(r)r^2{\rm d}r\propto a^{5/4}$
for the Bahcall-Wolf density profile $n_\star(r)\propto r^{-7/4}$.
For convenience, we write $R_\star(r < a)= C_\star a$, where $C_\star$ is the constant differential rate in the empty regime.
In this form,  the total disruption rate in the empty loss cone regime reads as
\be \label{eq:R_empty}
R_{\rm tot, empty} = C_\star a_{\rm full} \ ,
\ee 
for $a_{\rm full} < r_{\rm h}$ and $R_{\rm tot, empty}\approx C_\star r_{\rm h}$ for  $a_{\rm full} > r_{\rm h}$ (assuming that the stellar density decreases rapidly beyond $r_{\rm h}$).

In the full loss cone regime, stars follow a uniform distribution in $J^2\propto r_{\rm p}$ \citep{Cohn1978}. A fraction $2r_{\rm t}/a$ of stars  reside in the loss cone and get fully or partially disrupted within one orbital period, resulting in full TDEs or non-repeating pTDEs, respectively. 
The differential disruption rate in this regime is therefore
\be 
\frac{d}{dr} R_\star\Big|_{r=a} = \frac{1}{T_{\rm obt}(a)}  \frac{2r_{\rm t}}{a}  \frac{ d N_\star(r)}{dr} =  C_\star \left(\frac{a}{a_{\rm full}} \right)^{-9/4}   \ ,
\ee 
where we have used 
the continuity condition of the
differential disruption rate at $a=a_{\rm full}$ in the second  equal sign.
Consequently, we obtain the 
total disruption rate in the full loss cone regime as 
\be \label{eq:R_full}
R_{\rm tot, full} = \int_{a_{\rm full}}^{r_{\rm h}} \frac{dR_\star}{dr} \ dr = \frac{4}{5} C_\star a_{\rm full} \left[1- (a_{\rm full}/r_{\rm h})^{5/4} \right] \ ,
\ee 
for $a_{\rm full} < r_{\rm h}$ and $R_{\rm tot, full}=0$ for  $a_{\rm full} > r_{\rm h}$. 

In summary, the total disruption rate contributed by stars in both full and empty loss cone regimes is
\be 
R_{\rm tot} = R_{\rm tot, empty} + R_{\rm  tot, full} \approx C_\star r_{\rm h}\ ,
\ee 
where we have used Eqs.~(\ref{eq:full_empty}, \ref{eq:R_empty}, \ref{eq:R_full}) and the approximation sign is accurate to percent level for $M_\bullet>10^5 M_\odot$.
Specifically,  the full loss cone regime contributes  $\leq 4/9$ ($= 2/9$ “full TDEs” +$2/9$ “non-repeating pTDEs”  if the critical disruption radius for pTDEs is exactly $2r_{\rm t}$ ) of star disruptions, and the remaining $\geq 5/9$ disruptions (“rpTDEs”) are from the empty loss cone regime. Note that we count rpTDEs by the number of stars disrupted in a series of disruptions, instead of the number of observable flares produced.  In practice, only rpTDEs with short orbital period ($T_{\rm obt} \lesssim 10$ yr) can be identified as rpTDEs, which are the focus of this work.

The branch ratios obtained above are based on the Bahcall-Wolf  density profile $n_\star(r)\propto r^{-7/4}$, which holds only for relaxed nuclear stellar clusters.
For general nuclear stellar clusters with density profiles $n_\star(r)\propto r^{-\Gamma}$, a similar calculation shows that the full loss-cone regime contributes $\leq \frac{2}{2\Gamma+1}$ of stellar disruptions, and the empty regime contributes $\geq \frac{2\Gamma-1}{2\Gamma+1}$ (for $\Gamma > 0.5$). 

\subsection{Fractional rate of observable rpTDEs}

With the relations derived in the previous subsection,  it is straightforward to calculate the fraction of stars that end as rpTDEs (with period less than $T_{\rm obt}$)
among all star disruptions as 
\be\label{eq:f_rpTDE}
f_{<T_{\rm obt}}:=\frac{R_\star(< T_{\rm obt})}{R_{\rm tot}} \approx
\frac{a(T_{\rm obt})-a_{\rm GW}(r_{\rm p})}{r_{\rm h}},
\ee 
where $a(T_{\rm obt})$ is the semi-major axis of the star with orbital period $T_{\rm obt}$,
\be \label{eq:a_obt}
\frac{a(T_{\rm obt})}{M_\bullet} = 4.7\times 10^4\, T_{\rm obt, 10 yr}^{2/3} M_{\bullet, 6}^{-2/3}\ ,
\ee 
and $a_{\rm GW}(r_{\rm p})$ is given by Eq.~(\ref{eq:a_JGW}),
with $r_{\rm p}=r_{\rm t}/\beta_{\rm rpTDE}$ being the pericenter distance of the star experiencing a partial disruption. We adopt 
$\beta_{\rm rpTDE}=0.5$ as the fiducial value (see 
Section~\ref{subsec:boundary} for discussion).
Note that Eq.~(\ref{eq:f_rpTDE}) applies only for 
$a(T_{\rm obt})\ge a_{\rm GW}(r_{\rm p})$; otherwise $f_{<T_{\rm obt}}=0$
as the stars will end up as EMRIs. Setting 
$a(T_{\rm obt})= a_{\rm GW}(r_{\rm p})$, we find the maximum SMBH mass beyond which no rpTDE can happen:\footnote{As noted in the end of Section~\ref{sec:basics}, the boundary between stellar EMRIs and rpTDEs is pretty sharp. In rare cases, rpTDEs from stars entering the GW dominated regime with fine-tuning parameters may evade the $M_\bullet = M_{\bullet,\rm max}(T_{\rm obt})$ boundary. }
\be \label{eq:M_max}
M_{\bullet,\rm max}(T_{\rm obt}) = 1.6\times10^6 M_\odot\ T_{{\rm obt}, 10 \rm yr}^{4/9}\, \beta_{\rm rpTDE}^{-4/3} m_{\star,\odot}^{-4/9} R_{\star,\odot}^{4/3}\ .
\ee 
Similar to the GW radius $a_{\rm GW}$, the critical SMBH mass $M_{\bullet,\rm max}(T_{\rm obt})$ also has a weak dependence on the assumed stellar density profile. 
This relation defines a criterion of rpTDEs formed in the loss cone channel. 
In fact, Eq.~(\ref{eq:f_rpTDE}) can be written as
\be\label{eq:f_rpTDE2}
f_{<T_{\rm obt}}\approx 2.2\times 10^{-3}T_{\rm obt,10yr}^{2/3}M_{\bullet,6}^{-1/6} 
\left[1-\left({M_\bullet\over M_{\rm \bullet,max}}\right)^{3/2}\right]\ .
\ee
In deriving Eq.~(\ref{eq:f_rpTDE}), we have assumed the Bahcall-Wolf density profile, which is a good approximation for nuclear stellar clusters around SMBHs with $M_\bullet < M_{\bullet,\rm max}(T_{\rm obt})$. 

\begin{figure}
\centering
\includegraphics[width=0.5\textwidth]{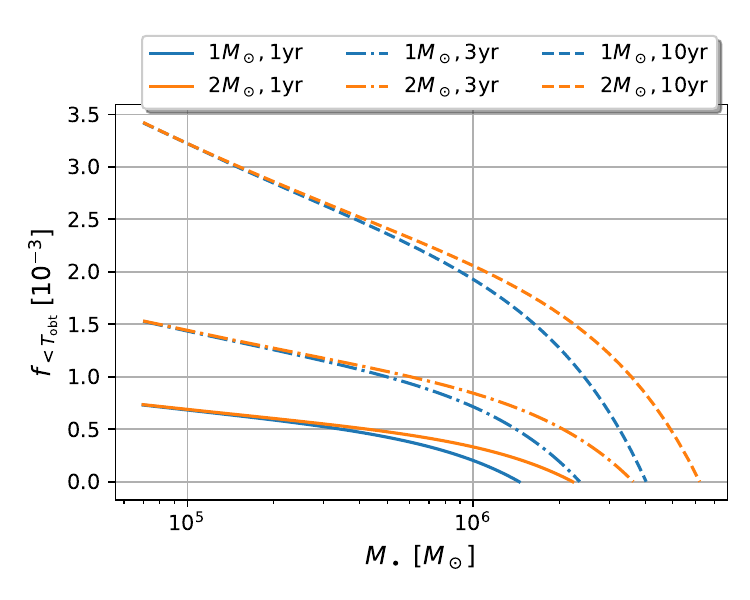}
\caption{Fraction of rpTDEs $f_{< T_{\rm obt}}(M_\bullet)$  among all star disruptions as a function of SMBH mass for $T_{\rm obt}\in\{1, 3, 10\}$ yr obtained from Eqs.~(\ref{eq:f_rpTDE}-\ref{eq:f_rpTDE2}), where we have used $\beta_{\rm rpTDE}=0.5$. The blue and orange lines represent
the fractional rates of rpTDEs from solar type stars ($1 M_\odot$) and more massive stars ($2 M_\odot$), respectively.}
\label{fig:f_rpTDE}
\end{figure}

In Fig.~\ref{fig:f_rpTDE}, we plot the expected rpTDE fraction $f_{<T_{\rm obt}}$ for $\beta_{\rm rpTDE}=0.5$ and  main sequence stars assuming a mass-radius relation $R_\star\propto m_\star^{0.8}$.  Clearly, $f_{< T_{\rm obt}}$ is larger for larger orbital period $T_{\rm obt}$ where a larger amount of stars are available to feed the loss cone. 
More massive main sequence stars allow for heavier SMBHs to generate rpTDEs as $M_{\bullet, \rm max} \propto m_\star^{0.62}$.

\begin{figure}
\centering
\includegraphics[width=0.5\textwidth]{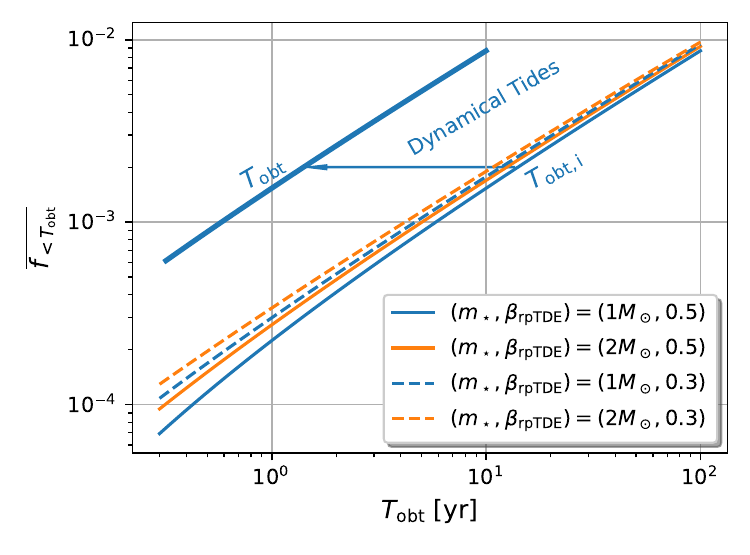}
\caption{Average rpTDE fraction in the loss cone channel is insensitive to the stellar type or the exact loss cone boundary $\beta_{\rm rpTDE}$.
The four “diagonal" lines are obtained from Eqs.~(\ref{eq:f_rpTDE2}, \ref{eq:f_avg}). In addition, dynamical tides may drive stars from larger “initial” orbital period $T_{\rm obt, i}$ to 
shorter orbital period $T_{\rm obt}$, where the stars repeatedly get disrupted and are observed. As a result, dynamical tides increase the rate of observable rpTDEs by a factor of a few (see Section~\ref{subsec:dynamical tides}). 
Here the heavy blue line is obtained from the light blue line by assuming $T_{\rm obt}=T_{\rm obt,i}/10$.}
\label{fig:f_avg}
\end{figure}

As a measure of relative occurrence rate of rpTDEs, we define an average fraction 
\be \label{eq:f_avg}
\overline{f_{< T_{\rm obt}}} :=\frac{\int_{M_{\bullet, \rm min}}^{M_{\bullet, \rm max}} f_{< T_{\rm obt}}(M_\bullet) R_{\rm tot}(M_\bullet) \frac{dN_\bullet}{dM_\bullet} \ dM_\bullet}{\int_{M_{\bullet, \rm min}}^{M_{\bullet, \rm max}} R_{\rm tot}(M_\bullet) \frac{dN_\bullet}{dM_\bullet}\ dM_\bullet}\ , 
\ee 
where $dN_\bullet/dM_\bullet$ is the SMBH mass function, and  $R_{\rm tot}$ is the total rate of both full TDEs and partial TDEs. Based on the TDEs observed in ZTF survey, \citet{Yao:2023rbr} infer the TDE rate as a function of SMBH mass, 
\be\label{eq:mass_func}
R_{\rm tot}(M_\bullet) \frac{dN_\bullet}{d M_\bullet} \propto M_\bullet^{-1.25} \ ,
\ee 
in the range of $(10^{5.3}, 10^{7.3}) M_\odot$. In Fig.~\ref{fig:f_avg}, we show the average rpTDE fraction for main sequence stars of different masses. 
We see the little dependence of $\overline{f_{< T_{\rm obt}}}$ on the stellar mass and $\beta_{\rm rpTDE}$
as expected from Fig.~\ref{fig:f_rpTDE} and Eq.~(\ref{eq:f_rpTDE}).

\begin{figure}
\centering
\includegraphics[width=0.5\textwidth]{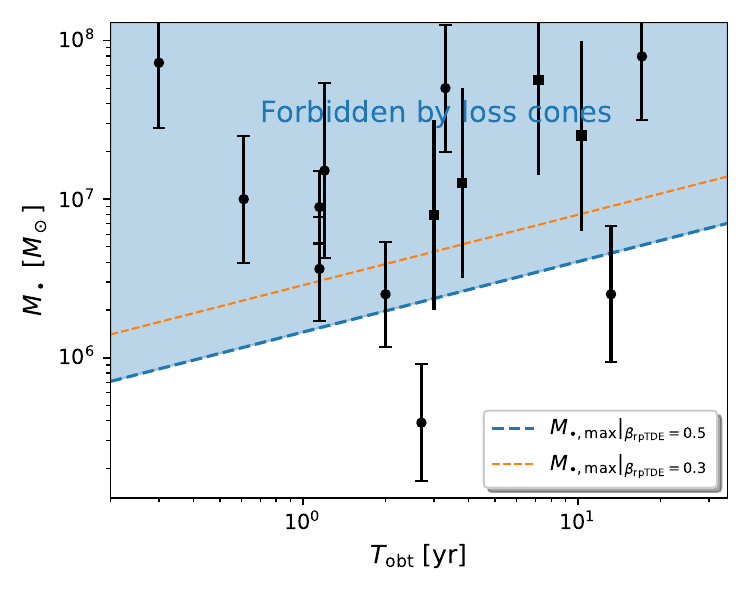}
\caption{Statistics of rpTDE candidates in the $T_{\rm obt}-M_\bullet$ space (see Table.~\ref{table}).
The shadowed region is forbidden by the loss cone channel, and the boundary is obtained from Eq.~(\ref{eq:M_max}) assuming solar type stars and different penetration factors of rpTDEs $\beta_{\rm rpTDE}$ .}
\label{fig:stat}
\end{figure}

In addition to the fraction of rpTDEs among all star disruptions $f_{< T_{\rm obt}}$,  Eq.~(\ref{eq:M_max}) provides a more direct criterion for examining formation channels of individual rpTDEs. In Fig.~\ref{fig:stat}, we plot the rpTDE candidates reported in the literature (see Table.~\ref{table}) in the $T_{\rm obt}-M_\bullet$ diagram. We find the majority of them violate the criterion $M_\bullet < M_{\bullet, \rm max}(T_{\rm obt})$ of the loss cone channel. This observation indicates most of these candidates are either not rpTDEs or they are rpTDEs from  alternative formation channels. In Section~\ref{sec:Hills}, we will examine the Hill mechanism as an alternative channel.

\subsection{Loss cone boundary of rpTDEs}\label{subsec:boundary}

The loss cone boundary of full TDEs is generally defined as $r_{\rm p}= r_{\rm t}$, i.e., a single pericenter passage within which ($r_{\rm p}\leq r_{\rm t}$) is expected to fully disrupt the star.
In a similar way, we can define the loss cone boundary $r_{\rm p}= \beta^{-1}_{\rm rpTDE} r_{\rm t}$  of rpTDEs such that a star with $r_{\rm p}\leq \beta^{-1}_{\rm rpTDE} r_{\rm t}$ is expected to be destroyed by a series of repeating partial disruptions. The criterion of rpTDEs is therefore that the cumulative timescale of repeating partial disruptions is shorter than the timeslcae of any processes that may change the stellar orbit. One of these processes is scatterings with other stars.

A number of previous studies have estimated the boundary or rpTDEs. 
Analytical argument of \cite{Coughlin2022} by equating the maximum
internal gravity of the star to the external tidal
field from the SMBH, and numerical simulations show that the “partial disruption radius”, where a star starts to lose mass from the envelope,
is $\beta\simeq 0.55-0.6$,
independent of star properties  \citep{Guillochon2013,Nixon:2021njx}, or $\beta\simeq 0.5-0.55$ depending on the equation of state \citep{Chen:2024amz}. Recent report of exponential growing flare energy in 5 consecutive flares of AT 2023uqm also shows good agreement with  simulations of rpTDEs with $\beta\simeq 0.55-0.6$ \citep{Wang:2025oyc}. 
Following the analyses above, we use the fiducial value 
\be \label{eq:beta_rpTDE}
\beta_{\rm rpTDE}=0.5\ ,
\ee as the loss-cone boundary of rpTDEs in this work.

In addition to the predictions based on analytical modeling or numerical simulations, long term observations of Sgr A$^\star$  spot a number of S stars (S2,
S4716, S62, S4714) on tight orbits with $r_{\rm p} = 120$, $100$,  $16$, $12$  AU \citep{Ghez2005,Gillessen2009,Peissker2020,Peissker2020b,Peissker2022}, respectively,
which also impose useful constraints on $\beta_{\rm rpTDE}$. Specifically, the stability of S4714 star with $m_\star\approx 2M_\odot$ \citep{Peissker2022} imposes a lower bound 
\be \label{eq:beta_S}
\beta_{\rm rpTDE} > 0.1\ ,
\ee which is  consistent with Eq.~(\ref{eq:beta_rpTDE}).

\subsection{Role of dynamical tides}\label{subsec:dynamical tides}

For a star on a highly eccentric orbit, oscillatory modes in the star are excited by the tidal force of the central SMBH during each pericenter passage.
Previous studies have examined the effects of such dynamical tides on the binary orbit \citep{Mardling1995,Lai:1997wh, Ivanov2004,Ivanov:2007we,Vick2018,Wu2018,Yu2021,Yu2022,Lau:2025cmv}.
In particular, depending on the pericenter distance and orbital period, there are 
two distinct long-term evolution tracks of the tidally-excited modes over multiple 
passages: regular quasi-steady oscillations (including larger-amplitude oscillations due to to a resonance between the orbital frequency and the mode frequency) and chaotic growth of oscillation modes.

In the regular oscillation regime, the mode energy $E_\alpha$ (where $\alpha$ is the mode
index) reaches a quasi-steady state (where the mode energy gained by each pericenter passage
is balanced by the energy dissipation over an orbital period), the timescale of orbital decay $t_{\rm tide}=|a/\dot a|_{\rm tide}$ induced by the tide depends on the mode damping time and is expected to be long \citep{Vick2018}. In the chaotic tide regime
(also called “diffusive tide", \cite{Wu2018}), the oscillation mode can grow to a large 
amplitude, followed by nonlinear dissipation, i.e., tidal heating.  Depending on 
the location of energy injection in the star, 
the star may be puffed up and become more susceptible to tidal stripping \citep{Lu2023,Linial2024p,Yao2025}. 
Multiple episodes of such chaotic mode
growth and dissipation cycle can lead to rapid and significant orbital decay \citep{Vick2019}.

The key condition for chaotic mode growth is \citep{Vick2018,Wu2018}
\be
\omega_\alpha |\Delta T_{\rm obt}|=\omega_\alpha T_{\rm obt}\,
\left({3\Delta E_\alpha\over 2|E_{\rm obt}|}\right)\gtrsim 1,
\label{eq:condition}\ee
where $\omega_\alpha$ is the mode frequency and $\Delta T_{\rm obt}$
is the change of the orbital 
period due to the “first" (i.e. when the initial mode energy is zero)
pericenter energy transfer $\Delta E_\alpha$ (where $E_{\rm obt}=-GMm_\star/2a$ is the orbital energy). Physically, $\omega_\alpha |\Delta T_{\rm obt}|$ is the phase change of the
oscillation mode due to the tidal energy transfer at pericenter. For a highly eccentric orbit, we can write
\be
\Delta E_\alpha={GM^2R_\star^5\over r_{\rm p}^6}\,{\cal E}(\beta),
\ee
where ${\cal E}$ is a dimensionless function 
(called $T_2$ by Press \& Teukolsky 1975) that depends on $\beta=r_{\rm t}/r_{\rm p}$, $\omega_\alpha/\Omega_{\rm p}$ (where $\Omega_{\rm p}=\sqrt{GM/r_{\rm p}^3}$ is the pericenter orbital 
frequency) and $Q_\alpha$ (the dimensionless mode overlap integral).
Thus 
\be
\omega_\alpha |\Delta T_{\rm obt}|=5.9\times 10^4\hat\omega_\alpha\, \beta^6\,
{\cal E}(\beta)\,
{m_{\star\odot}^{3/2}\over R_{\star\odot}^{5/2}M_{\bullet 6}^{2/3}}\left({T_{\rm obt}\over 10\,{\rm yrs}}\right)^{5/3},
\ee
where $\hat\omega_\alpha=\omega_\alpha (Gm_\star/R_\star^3)^{-1/2}$.
Using the approximate expression for ${\cal E}(\beta)$ given by \cite{Lai:1997wh} (see \cite{Vick2019}), valid for $\omega_\alpha/\Omega_{\rm p}$ larger than a few, we have
\begin{eqnarray}
&& \omega_\alpha |\Delta T_{\rm obt}|=8.8\times 10^5
\,\hat\omega_\alpha^4\,Q_\alpha^2\, \beta^{-3/2}
{m_{\star\odot}^{3/2}\over R_{\star\odot}^{5/2}M_{\bullet 6}^{2/3}}
\left({T_{\rm obt}\over 10\,{\rm yrs}}\right)^{5/3}\nonumber\\
&&\qquad\qquad\quad \times \exp(-4\sqrt{2}\,\hat\omega_\alpha/\beta^{3/2}).
\label{eq:omegaT}\end{eqnarray}

As an estimate, we model the star as a $\Gamma=5/3$ polytrope, with the f-mode
dominating the tidal interaction, and $\hat\omega_\alpha=1/46$, $Q_\alpha=0.49$ 
\citep{Vick2019}. Equation~(\ref{eq:omegaT}) then gives $\omega_\alpha|\Delta T_{\rm obt}|
\simeq 2\times 10^{-4}$ and 2 for $\beta=0.5$ and $2/3$, respectively 
(with $T_{\rm obt}=10$~years and other fiducial numbers given). We thus conclude
that chaotic tide can operate in the pre-rpTDE binaries with $\beta\gtrsim 0.5$ and $T_{\rm obt}\gtrsim 10$~years\footnote{In addition, two effects can make the system more likely to develope or stay in the chaotic tide regime: (i) The star can have an appreciable rotation, which may change the mode frequency and significantly increase $\Delta E_\alpha$ (e.g.
when the star has a retro-grade rotation relative to the orbit); (ii) In the presence of 
a “pre-existing" oscillation energy $E_{\rm resid}$ (such as the residual 
mode energy left over after a previous episode of chaotic mode growth and damping; see Vick et al.~2019), the condition (\ref{eq:condition}) for chaotic mode growth is replaced by
$\omega_\alpha T_{\rm obt} (3\Delta E_{\rm \alpha,max}/2|E_{\rm obt}|)\gtrsim 1$, with
$\Delta E_{\rm \alpha,max}\simeq \Delta E_\alpha+2\sqrt{E_{\rm resid}\Delta E_\alpha} $ \citep{Wu2018,Vick2019}. For $E_{\rm resid}>\Delta E_\alpha$, chaotic mode growth can be
achieved more easily than what Eq.~(\ref{eq:omegaT}) indicates.}.

We can now outline and estimate how chaotic tides may influence the evolution of a SMBH-star binary  prior to rpTDE. Consider a star scattered into a highly eccentric orbit
with $r_{\rm p}\simeq 2r_{\rm t}$ (and period $T_{\rm obt}\gtrsim 10$~years), sufficient 
for chaotic tide to operate. The tidally excited stellar oscillation mode would grow chaotically (or “diffusively", with the mode energy increasing approximately linear in time on average), followed by rapid mode decay. The system may go through multiple episodes of such mode growth/damping cycles \citep{Vick2019}; 
the net effect of these is the orbital shrinkage on the timescale 
\be
t_{\rm tide}\sim \left({|E_{\rm obt}|\over \Delta E_\alpha}\right)\, T_{\rm obt}
\sim {T_{\rm obt}^2\over |\Delta T_{\rm obt}|},
\ee
where $\Delta E_\alpha$ and $\Delta T_{\rm obt}$ as the same as in Eq.~(\ref{eq:condition}).
Thus
\be
{t_{\rm tide}\over T_{\rm obt}}\sim
{2.3\times 10^5\over \omega_\alpha|\Delta T_{\rm obt}|}
\,\,\hat\omega_\alpha \left({T_{\rm obt}\over 10~{\rm yr}}\right)
\propto T_{\rm obt}^{-2/3}.
\ee
On the other hand, the timescale for the angular momentum relaxation due to scatterings
(see Eq.~\ref{eq:tj}) is given by 
\be
{t_J\over T_{\rm obt}}\sim 2\times 10^5\,\beta^{-1}M_{\bullet 6}^{29/24}m_{\star\odot}^{-4/3}R_{\star\odot}
\left({T_{\rm obt}\over 10~{\rm yr}}\right)^{-3/2}
\ee
The fact that $t_{\rm tide}$ is comparable to $t_J$ for $T_{\rm obt}\sim 10$~years indicates
that there is a significant chance that appreciable orbital decay driven by chaotic tide
can happen before rpTDE occurs, although there is an interplay between scatterings and 
tide-driven orbital evolution -- we plan to investigate this interplay and other related issues  (e.g., stellar structure change due to tidal heating) in a future study.

Overall, our study in this section suggests that in the lose-cone scenario,
many rpTDEs with observed period $T_{\rm obt}$ may be “captured" at a larger (“initial") period $T_{\rm obt,i}$. While more quantitative calculations would be needed to determine
$T_{\rm obt,i}$, previous works (e.g. \cite{Vick2019}) in a different context suggests
that $T_{\rm obt,i}\sim 10 T_{\rm obt}$ is possible. With this consideration, 
the rpTDE fraction for a given $T_{\rm obt}$ can be increased by a factor of a few
(see Fig.~\ref{fig:f_avg}). 

Note that we have focused on the role of dynamical tides in reducing the stellar orbital size $a$, 
while ignored the response of the stellar structure to energy dissipation. 
With this simplified assumption, the critical value $M_{\bullet, {\rm max}}(T_{\rm obt})$, determined by $a(T_{\rm obt})= a_{\rm GW}(r_{\rm p})$, is not affected by dynamical tides.

\section{Hills mechanism}\label{sec:Hills}

Hills mechanism is another commonly invoked channel depositing stars onto extremely eccentric orbits around SMBHs by tidally disrupting stellar binaries \citep{Hills1988, Miller2005}. 
After a binary disruption, one star is captured by the SMBH and  the other is in general ejected from the SMBH. This is a possible source of rpTDEs and also a likely origin of hypervelocity stars (HVSs) found in the Milky way and local galaxies \citep{Yu2003, Gould2003, Brown:2005ta, Koposov2020}.
In this section, we will analyze whether the predictions of the Hills mechanism are compatible with observations of both rpTDEs and HVSs.

\subsection{Basics}

Consider a stellar binary with two equal-mass components (each $m_\star$) and an initial binary separation $a_{\rm b}$ orbiting around a SMBH with “initial"
semi-major axis $a_{\rm in}$.
The binary is disrupted when the pericenter distance is about the binary tidal radius
\be
r_{\rm t, b}= \left({M_\bullet\over 2m_\star}\right)^{1/3} a_{\rm b}.
\ee
The general outcome of the binary disruption is that 
one star gains some energy and gets ejected, the other loses approximately the same amount of energy and gets captured by the SMBH. The captured one is a potential source for full TDEs and partial TDEs.
The pericenter distance of the captured star is $r_{\rm p} \simeq r_{\rm t,b}$, and its
(specific) orbital binding energy can be written as
\be E_{\rm bind}={\zeta_E\over 2}{GM_\bullet a_{\rm b}\over {r_{\rm t,b}}^2},
\ee 
Where $\zeta_E\sim 1 $ is a numerical factor that depends on the orbital phase 
at binary disruption \citep{Sari2010}. For circular orbit and $r_{\rm p}\simeq r_{\rm t,b}$,
the value of $\zeta_E$ is distributed in the range $[0.5,3.4]$ with $\zeta_E\simeq 2.6$ at the peak of the distribution \citep{Yu2024}.
Thus the semi-major axis and eccentricity of the captured star are
\be \label{eq:a_capture}
a={\zeta_E^{-1}} \frac{r_{\rm t,b}^2}{a_{\rm b}}={\zeta_E^{-1}} \left(\frac{M_\bullet}{2m_\star} \right)^{2/3}  a_{\rm b}\ ,
\ee 
and 
\be \label{eq:e_capture}
1-e = \frac{r_{\rm p}}{a} \simeq \zeta_E \frac{a_{\rm b}}{r_{\rm t,b}} =  \zeta_E \left(\frac{2m_\star}{M_\bullet}\right)^{1/3}.
\ee 
The corresponding orbital period is 
\be \label{eq:Tobt_a}
T_{\rm obt} = 0.63 \ \zeta_E^{-3/2} M_{\bullet, 6}^{1/2} \left(\frac{a_{\rm b}}{2.5 R_\star}\right)^{3/2} R_{\star, \odot}^{3/2} m_{\star,\odot}^{-1}\ {\rm yr} \ .
\ee  
For main sequence stars with $R_\star \propto m_\star^{0.8}$, the orbital period  $T_{\rm obt}(a_{\rm b} = 2.5 R_\star)$ of the captured star has a weak dependence on the main sequence star mass $\propto m_\star^{0.2}$.
Note that the initial binary separation $a_{\rm b}$ must satisfy
\be \label{eq:a_bound}
2.5 R_\star \lesssim a_{\rm b} \lesssim a_{\rm b, hard}
\approx {2 Gm_\star\over \sigma^2(r)},
\ee
where $\sigma(r)$ is the local velocity dispersion of stars at $r\sim a_{\rm in}$.
In Eq.~(\ref{eq:a_bound}), the first inequality comes from the Roche stability 
criterion and the second comes from the fact that we expect the binary be “hard" to avoid being dissolved in the star cluster.
In the following analysis we simply use $\sigma(r)=\sigma_\star$ at the radius of influence ($r_{\rm h}$), where most star disruptions originate.
The pericenter distance and the semimajor axis of the capture star fall in the range of 
    \be \label{eq:rp_range}
    \begin{aligned}
 2r_{\rm t}\approx  2.5\left(\frac{M_\bullet}{2m_\star} \right)^{\!1/3}\!R_\star &\lesssim r_{\rm p} \lesssim \left(\frac{M_\bullet}{2m_\star} \right)^{\!1/3} a_{\rm b, hard}\ , \\ 
2.5{\zeta_E^{-1}}\left(\frac{M_\bullet}{2m_\star} \right)^{\!2/3}\!R_\star & \lesssim a \lesssim {\zeta_E^{-1}}\left(\frac{M_\bullet}{2m_\star} \right)^{\!2/3} a_{\rm b, hard}\ .
\end{aligned}
\ee 

\begin{figure}
\centering
\includegraphics[width=0.5\textwidth]{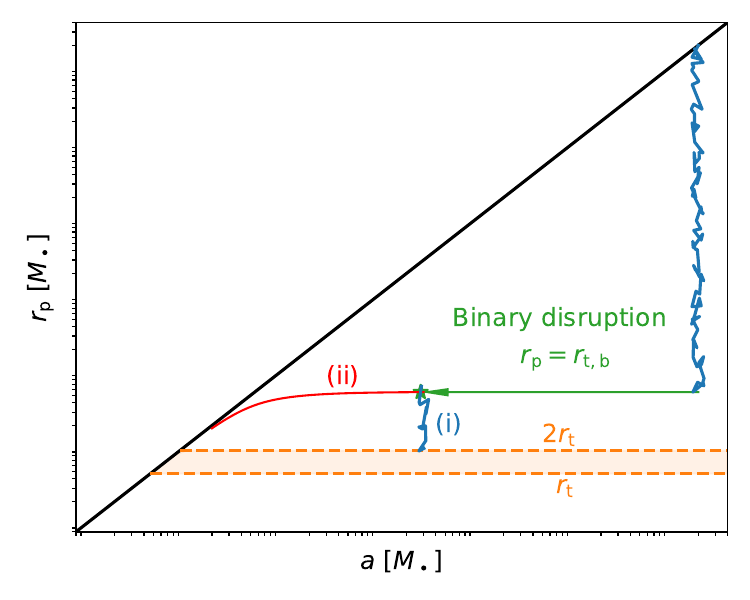}
\caption{
A cartoon illustrating the outcomes of binary disruption, leading to rpTDEs and stellar EMRI. In the 1st stage, a stellar binary 
is randomly scattered by background stars and gets disrupted when the pericenter distance $r_{\rm p}$ approaches $r_{\rm t,b}$.  
Immediately after the binary disruption, the captured star has a pericenter distance larger than the stellar tidal radius $r_{\rm t}$.
In the 2nd stage, the captured star is either (i) randomly scattered by background stars (when $t_J\lesssim t_{\rm GW}$) and finally ends as a rpTDE when it diffuses to the loss cone boundary of rpTDEs $r_{\rm p}\simeq 2r_{\rm t}$, or (ii) driven by gravitational radiation (when $t_{\rm GW}\lesssim t_{J}$) to become a stellar EMRI.}
\label{fig:Hills_cartoon}
\end{figure}
The captured star can follow two different evolutionary tracks as described in Section \ref{sec:loss cone}, depending on the ratio $t_J/t_{\rm GW}$, which in terms depends on $a,~e$ and other parameters 
(see Fig.~\ref{fig:Hills_cartoon}): For $t_J/t_{\rm GW}\gtrsim 1$, 
it is primarily driven by gravitational radiation to become a stellar EMRI;
for $t_J/t_{\rm GW}\lesssim 1$, it is driven by scatterings to become a rpTDE.
In the latter case, the rpTDE would have a period given by Eq.~(\ref{eq:Tobt_a}).
We note that some observed rpTDEs have a rather short period (e.g. 114 days for
ASASSN-14ko, \cite{Payne:2020tfp}); if they are formed in the Hill mechanism, very tight initial binaries
(with $a_{\rm b}$ a few times $R_\star$) would be required (cf. \cite{Cufari:2022szx}).

\begin{figure}
\centering
\includegraphics[width=0.5\textwidth]{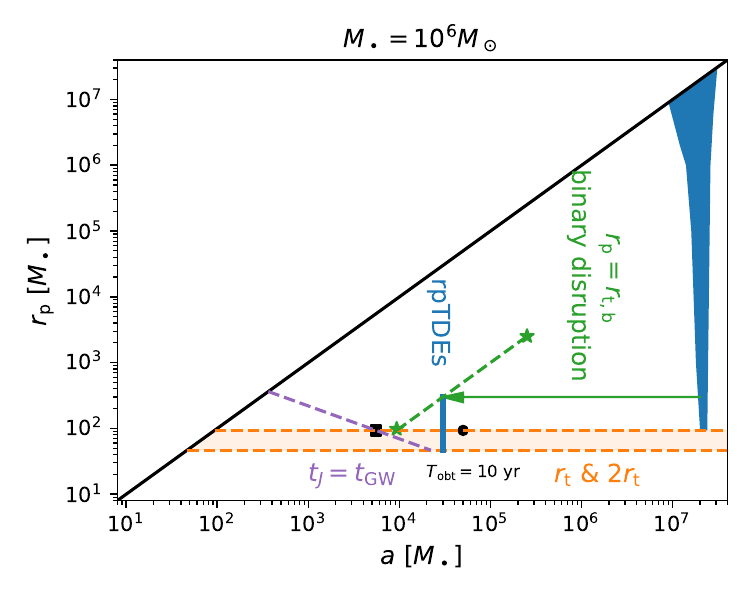}\\
\includegraphics[width=0.5\textwidth]{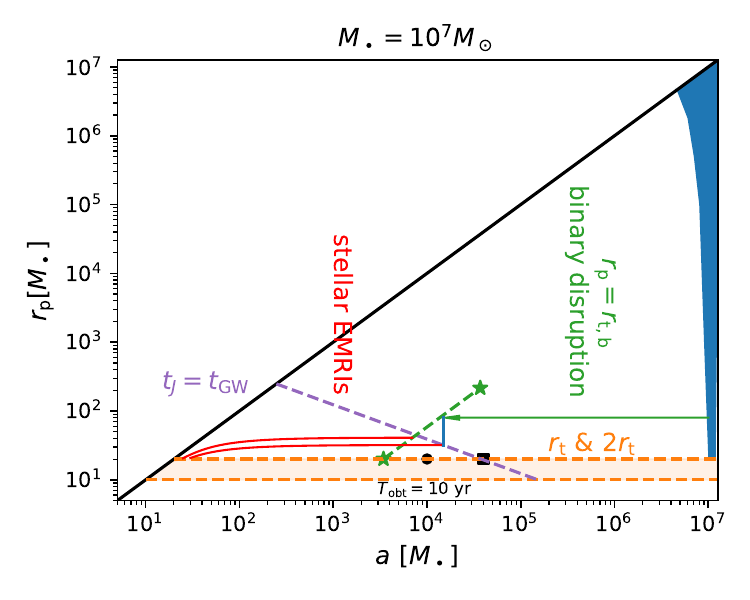}
\caption{Schematic plot of binary disruption and the subsequent evolution of the captured star. 
A binary is disrupted by the central SMBH when the binary approaches $r_{\rm t, b}$, with one of the two stars being ejected and the other being captured. 
In the $r_{\rm p}-a$ phase space, the captured star is deposited at the green dashed line, where the two $\star$ symbols denote the upper and the lower limits [Eq.~(\ref{eq:rp_range})], respectively. Depending on the initial separation of the binary, the captured star faces two possible fates: rpTDE or stellar EMRI. In the upper panel (for $M_\bullet=10^6 M_\odot$), the captured stars are in the 2-body scattering dominated regime [Eq.~(\ref{eq:Tobt_a})], where those with $a_{\rm GW} < a < a_{\rm 10\ yr}$ finally end as observable rpTDEs. In the lower panel ($M_\bullet=10^7 M_\odot$), the captured stars with $T_{\rm obt} < 10$ yr are in the GW emission dominated regime ($a_{\rm 10\ yr} < a_{\rm GW}$), where those with $r_{\rm p} \approx 2r_{\rm t}$ start the rpTDE phase immediately after the binary disruption, while others gradually circularize and end as stellar EMRIs.}
\label{fig:Hills}
\end{figure}

Similar to Fig.~\ref{fig:phase_diagram}, we analyze the outcomes of the Hills mechanism in the $r_{\rm p}-a$ space in Fig.~\ref{fig:Hills} (similar discussions can be found in earlier studies, e.g., \cite{Lu2023,Linial:2022bjg}).
In the upper panel (for $M_\bullet=10^6 M_\odot$), a binary is randomly scattered by background stars in the nuclear stellar cluster and gets disrupted by the central SMBH when its pericenter distance approaching $r_{\rm t,b}$. The binary disruption leads to  one of the stars to be ejected and the other captured by the SMBH.
The captured star is deposited in the 2-body scattering dominated regime.
Depending on the initial separation of the stellar binaries,
the captured stars with $r_{\rm p}\approx  2 r_{\rm t}$ immediately enter the rpTDE phase, while those with $r_{\rm p} > 2r_{\rm t}$ will first experience random 2-body scatterings until being kicked onto a low-angular momentum orbit with $r_{\rm p}\approx 2r_{\rm t}$ and end as rpTDEs (see also Fig.~\ref{fig:Hills_cartoon}), among which only those with $a < a_{\rm 10\ yr}$ are recognized as observable rpTDEs 
and the remaining ones with longer orbital periods can be easily identified as non-repeating TDEs
\citep{Zhong2022,Bortolas:2023tlq,Broggi:2024qkj}.

In the lower panel of Fig.~\ref{fig:Hills} (for 
$M_\bullet=10^7 M_\odot$), where the critical point $(a=a_{10\ \rm yr},  r_{\rm p}=2r_{\rm t})$ is located in the GW emission dominated regime, i.e., $ a_{10\ \rm yr} < a_{\rm GW}(r_{\rm p})$. As a result,
captured stars with $r_{\rm p} \approx 2 r_{\rm t}$ immediately end as observable rpTDEs.
The remaining ones with $r_{\rm p} > 2 r_{\rm t}$ and $a < a_{\rm 10 \ yr}$ gradually circularize due to GW emission and become stellar EMRIs. 

To summarize, a critical value of the SMBH mass in the Hills mechanism is $M_{\bullet, \rm max}(T_{\rm obt})$ [Eq.~(\ref{eq:M_max})] at which $a_{\rm GW}(r_{\rm p}=2r_{\rm t}) = a (T_{\rm obt})$, similar to in the single-star loss cone channel. For lighter SMBHs disrupting stellar binaries, all captured stars with $a_{\rm GW} \leq a \leq a_{10\ \rm yr}$ end as observable rpTDEs (see upper panel of Fig.~\ref{fig:Hills}). For heavier SMBHs, observable rpTDEs are sourced by captured stars with $r_{\rm p} \approx 2 r_{\rm t}$ only. These captured stars are from disruptions of near-contact binaries with $a_{\rm b} \approx 2.5 R_\star$.

\subsection{Fractional rate of observable rpTDEs}

With the basic picture of binary disruption outlined in the previous subsection, we can calculate the rate of rpTDEs contributed by the Hills mechanism.
Let $f_{\rm b, hard}$ be the fraction of hard binaries in the
cluster. Since the relaxation timescale  $t_{{\rm b}, J}$ of stellar binaries in the angular momentum direction is comparable to that of single stars $t_J$,
hard binaries are expected to contribute a fraction of about $(1/2)f_{\rm b, hard}$ of stellar disruptions. To be classified as observable rpTDEs, additional requirements on orbital parameters $(a, r_{\rm p})$ of captured stars are needed, as discussed in the previous subsection:
\begin{equation}
    \begin{cases}
      a_{\rm GW}(r_{\rm p}=2r_{\rm t}) \leq a \leq  a(T_{\rm obt}) & \text{if}\ M_\bullet \leq M_{\bullet, \rm max}(T_{\rm obt})\\
      r_{\rm p}\approx 2r_{\rm t} \ \text{or equivalently} \  a_{\rm b}\approx 2.5 R_\star  & \text{if}\ M_\bullet > M_{\bullet, \rm max}(T_{\rm obt})\ .
    \end{cases}       
\end{equation}

\begin{figure}
\centering
\includegraphics[width=0.5\textwidth]{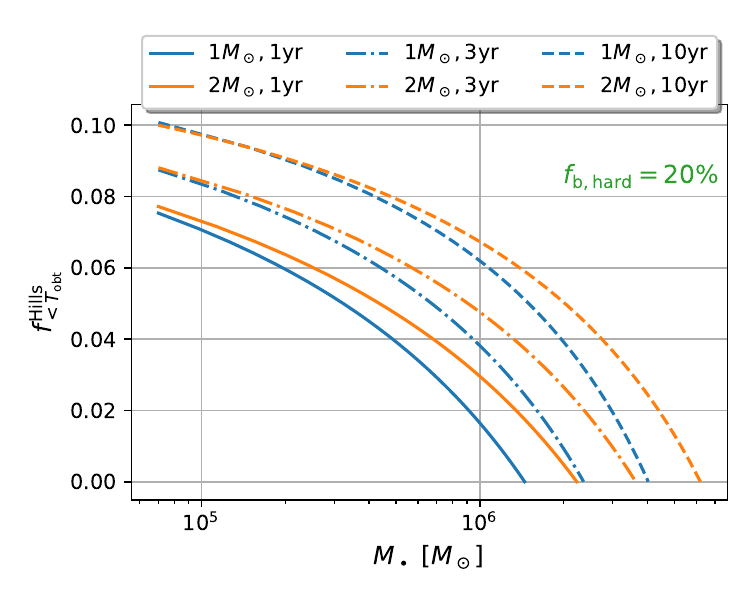}
\caption{Fraction of rpTDEs $f_{< T_{\rm obt}}^{\rm Hills}(M_\bullet)$ as a function of SMBH mass for $T_{\rm obt}\in\{1, 3, 10\}$ yr contributed by binary disruptions and calculated from Eq.~(\ref{eq:f_rpTDE_Hills}), assuming that a fraction $f_{\rm b, hard}=20\%$ of stars are in hard binaries.}
\label{fig:f_rpTDE_Hills}
\end{figure}

The rate of rpTDEs in the Hills channel clearly depends on the distribution of $a_{\rm b}$.
Assuming $\log(a_{\rm b})$ follows a uniform distribution in the range defined in Eq.~(\ref{eq:a_bound}), we find that the fractional rate  of rpTDEs (with period less than $T_{\rm obt}$) contributed by the Hills mechanism is 
\be \label{eq:f_rpTDE_Hills}
f_{< T_{\rm obt}}^{\rm Hills} \approx \frac{f_{\rm b, hard}}{2} \frac{\log\left(\frac{a(T_{\rm obt})}{a_{\rm GW}(r_{\rm p}=2r_{\rm t})}\right)}{\log\left(\frac{a_{\rm b, hard}}{2.5R_\star}\right)}\ ,
\ee 
where the factor $f_{\rm b, hard}/2$ is the fraction of star disruptions contributed by the Hills mechanism and the factor $\log(.)/\log(.)$ account for observable rpTDEs only.
In Fig.~\ref{fig:f_rpTDE_Hills}, we plot the expected rpTDE fraction $f_{<T_{\rm obt}}^{\rm Hills}$ for $\beta_{\rm rpTDE}=0.5$ and  main sequence stars assuming a mass-radius relation $R_\star\propto m_\star^{0.8}$. Similar to the loss cone channel, no rpTDE can happen for $M_\bullet > M_{\bullet, \rm max}$ in the Hills channel  \emph{if} the disrupted stellar binaries  are normal detached binaries.
Compared to Fig.~\ref{fig:f_rpTDE}, the Hills channel may dominate over the loss cone channel in producing observable rpTDEs if a fraction $f_{\rm b, hard}>20\%$ of stars are in hard binaries.

\begin{figure}
\centering
\includegraphics[width=0.5\textwidth]{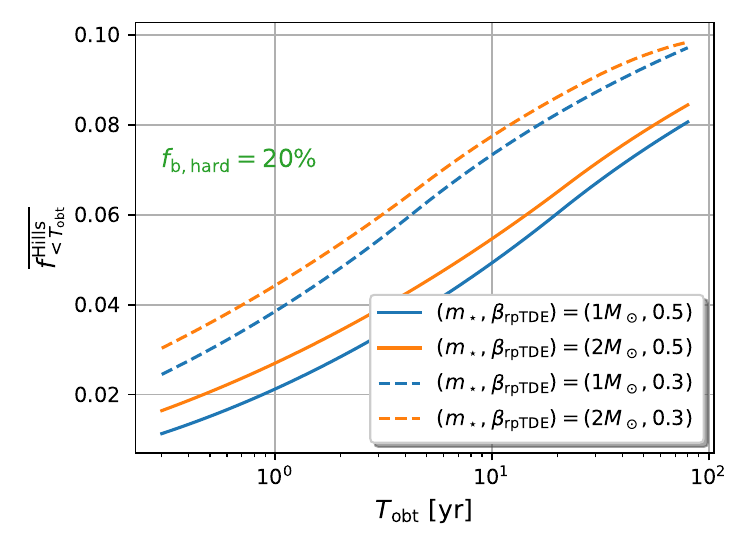}
\caption{Average rpTDE fraction contributed by the Hills channel assuming the binary separation $a_{\rm b}$ satisfies a log-uniform distribution.}
\label{fig:f_avg_Hills}
\end{figure}

Similar to Eq.~(\ref{eq:f_avg}) for the loss cone channel, we define an average fraction $\overline{f_{< T_{\rm obt}}^{\rm Hills}}$ for the Hills channel. As shown in Fig.~\ref{fig:f_avg_Hills}, $\overline{f_{< T_{\rm obt}}^{\rm Hills}}$ has a logarithmic dependence on the orbital period $T_{\rm obt}$, and as high as $\approx 5\%$ star disruptions are observable rpTDEs if $20\%$ of stars are in hard binaries and $\log(a_{\rm b})$ follows a uniform distribution.

\begin{figure}
\centering
\includegraphics[width=0.5\textwidth]{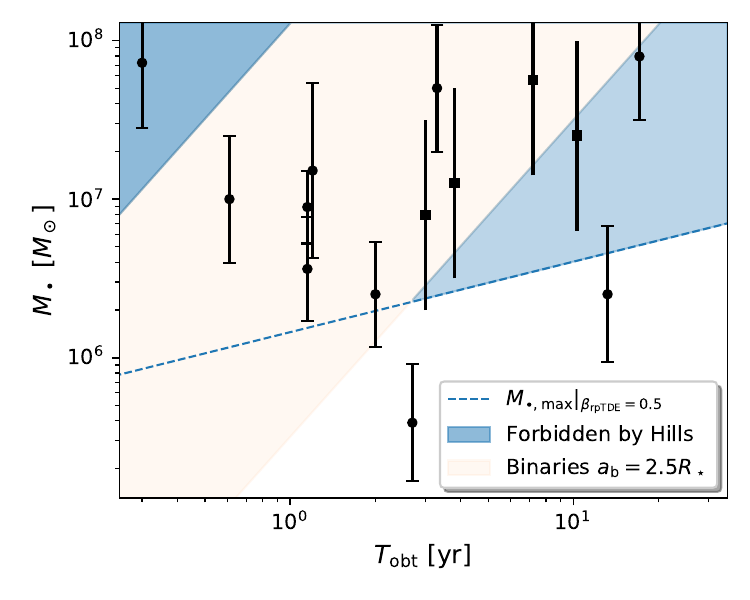}
\caption{Statistics of rpTDE candidates in the $T_{\rm obt}-M_\bullet$ space (see Table.~\ref{table}).
The blue dashed line is where $M_\bullet = M_{\bullet, {\rm max}}(T_{\rm obt})$ [Eq.~(\ref{eq:M_max})], beyond which no rpTDE can happen in the loss cone channel.
In the Hills channel, near contact binaries with $a_{\rm b}\approx 2.5 R_\star$ can evade this limit [Eq.~(\ref{eq:Tobt_a})], as shown in the orange shadowed region. 
The white region is contributed by detached binaries with $a_{\rm b} > 2.5 R_\star$.}
\label{fig:stat_Hills}
\end{figure}

In addition to the fractional rate of rpTDEs, the distribution of rpTDEs in the $T_{\rm obt}-M_\bullet$ diagram is another probe to the Hills channel. 
In Fig.~\ref{fig:stat_Hills}, we show the allowed and forbidden regions of the Hills mechanism in $T_{\rm obt}-M_\bullet$ space. 
The orange shadowed region where the majority of rpTDEs reside is sourced by the disruptions of near-contact binaries with $a_{\rm b}\approx 2.5 R_\star$. 
The white region where only two rpTDEs reside is contributed by normal detached binaries with $a_{\rm b} > 2.5R_\star$. The blue shadowed region is forbidden by the Hills mechanism.\footnote{The short-period rpTDE candidate ASASSN-14ko, with $T_{\rm obt}= 0.3$ yr and $\log_{10}(M_\bullet/M_\odot)=7.86^{+0.31}_{-0.41}$, lies in the forbidden region of the Hills mechanism (upper left corner of Fig.~\ref{fig:stat_Hills}). 
To explain the origin of this event with the Hills mechanism requires a contact star+compact object binary with a low binary separation $a_{\rm b} = 1 R_\odot $ and a low SMBH mass $M_\bullet=10^7 M_\odot$ (see \cite{Cufari:2022szx}).}
From Fig.~\ref{fig:stat_Hills}, 
we conclude that if the Hills mechanism dominates the production of rpTDEs, most of the rpTDE events are from near-contact binaries.

\subsection{Additional constraint: HVSs}

Following a binary disruption by a SMBH, a star is captured and possibly ends as a rpTDE as discussed in the previous subsection, and the other is ejected. This is likely the origin of HVSs found in the Milky way \citep{Hills1988,Yu2003,Brown:2005ta,Brown2014, Brown2018, Hirsch2005,Zheng2014, Koposov2020, Marchetti2022, Verberne:2024fag}.
The ejection velocity is given by (see Eq.~\ref{eq:a_capture})
\be \label{eq:v_ej}
\begin{aligned}
    v_{\rm ej} &= \sqrt{\frac{GM_\bullet}{a}} = \zeta_E^{1/2}\sqrt{\frac{ 2Gm_\star}{a_{\rm b}}} \left(\frac{M_\bullet}{2m_\star}\right)^{1/6} \\ 
&=3.6\times 10^3 \zeta_E^{1/2} \left(\frac{a_{\rm b}}{2.5R_\star}\right)^{-1/2} R_{\star, \odot}^{-1/2} m_{\star,\odot}^{1/3} M_{\bullet,6}^{1/6}\ {\rm km\ s}^{-1}\
.
\end{aligned}
\ee
For main sequence stars with $R_\star \propto m_\star^{0.8}$, 
the dependence of the ejection velocity for near-contact binaries on the stellar mass is weak with $v_{\rm ej}\propto m_\star^{-0.07}$.

In the previous subsection, we have shown that if the observed rpTDEs are produced by the Hills mechanism, then they likely have arisen from near-contact binaries.  From the same process, HVSs with $v_{\rm ej} \approx 4.5  \times 10^3 \zeta_E^{1/2}\ {\rm km\ s}^{-1}$ are expected to be ejected.
Assuming the Milky Way is a typical galaxy, its SMBH tidally disrupts stars at a rate of 
  $\approx 3\times 10^{-5}\ {\rm yr}^{-1}$. Roughly $10\%$ of these events are rpTDEs resulting from the Hills mechanism. Based on this, we expect approximately 100 HVSs ejected from Sgr A$^\star$ to exist within 100 kpc of Sgr A$^\star$. 

Confirmed and candidate HVSs are summarized in Table~\ref{tab:gc_hvs}, though the origin of these HVSs is still uncertain \citep{Han2025hvs}. Early-type stars are preferentially detected due to selection bias.
If a star were ejected from the Galactic Center (GC),  it must have decelerated from its initial velocity $v_{\rm ej}$ to its current observed value $v_{\rm GC}$ as it climbed out of the galactic potential well (see e.g. Fig.~4 of \cite{Brown2014}).
After correcting the difference in potential, we find that the ejection velocities of these HVS (candidates) are $\lesssim 10^3 \ {\rm km \ s}^{-1}$, except $v_{\rm ej}({\rm S5-HVS\ 1})\approx 2\times 10^3 \ {\rm km \ s}^{-1}$. 
Thus, these HVs did not originate from near-contact binaries,
but rather from binaries with $a_b\gg 2.5R_\star$.

In summary, the Hills mechanism has been proposed as a promising channel for sourcing both rpTDEs and HVSs. With this assumption, our analysis of rpTDEs in the $T_{\rm obt}-M_\bullet$ diagram shows that the majority of rpTDEs must be sourced by captured stars following SMBHs disrupting near-contact stellar binaries  (see Fig.~\ref{fig:stat_Hills}). 
Accompanying these rpTDEs,  HVSs are inevitably ejected with extremely high velocities [Eq.~(\ref{eq:v_ej})] in the same rate. 
If the Milky Way is a typical galaxy that produces TDEs (and rpTDEs), 
a large number of such HVSs moving away from Sgr A$^\star$ are expected.
However, HVSs found in the Milky Way have lower ejection velocities, therefore are the result of the SMBH disrupting normal detached binaries.
Due to this discrepancy, it remains unclear whether the Hills mechanism is the dominant channel for producing rpTDEs. 
More complete search of HVSs moving with velocities as high as $\sim 4\times 10^3$ km s$^{-1}$ will be critical for answering this question.

\section{Discussion and Conclusions}\label{sec:conclusion}

We have analyzed the predictions of the (single-star) loss cone channel and the Hills mechanism (binary tidal breakup) in producing rpTDEs, and compare them with candidate rpTDEs reported in the literature. We find that the loss cone channel generally predicts too low a rate of rpTDEs and the criterion $M_\bullet \lesssim 4\times 10^6 M_\odot (T_{\rm obt}/10\ {\rm yr})^{4/9}$
(a necessary condition for producing rpTDEs) is 
violated by the majority of the reported rpTDEs (see Fig.~\ref{fig:stat}). Dynamical tides may increase the rate of observable rpTDE by a factor of a few by driving stars at larger initial orbital period $T_{\rm obt, i}$ to shorter orbital period $T_{\rm obt}$, where the stars end as rpTDEs. The predicted fractional rate of rpTDEs is still lower than the observation by a factor of $\mathcal{O}(10)$ (see Fig.~\ref{fig:f_avg}).
Therefore, we arrive at the conclusion that the loss cone channel is not likely the dominant source of rpTDEs if the candidates reported are genuine rpTDEs. 

A similar analysis shows that the Hills mechanism can produce sufficient observable rpTDEs if a substantially fraction of stars ($f_{\rm b}\gtrsim 20\%$) in nuclear star clusters are in binaries (see Fig.~\ref{fig:f_avg_Hills}) and the predicted distribution of rpTDEs in $T_{\rm obt}-M_\bullet$ diagram is also compatible with the observed if the majority of disrupted stellar binaries by SMBHs are nearly contact (see Fig.~\ref{fig:stat_Hills}). This seems unnatural for stellar binaries. 
However, several recent theoretical studies suggest tight stellar binaries as a natural result of dynamical interactions with the central SMBH or flying-by stars in the cluster. For example,
\cite{Lu2023} and \cite{HTHuang2025} found that repeated tidal encounters with SMBH may perturb a stellar binary to high eccentricity, 
and harden the binary to separation $a_{\rm b}\lesssim 10 R_\star$. \cite{Winter-Granic:2023pei} showed that tidal perturbations from flying-by field stars work in a similar way in tightening stellar binaries in nuclear stellar clusters. \cite{Dodici2025} further predicted a population of 
close 
binaries in nuclear stellar clusters, though earlier studies (e.g., \cite{Bradnick2017}) predicted the majority stellar binaries merge during the orbital eccentricity excitation phase. 

Another inevitable consequence of disrupting near-contact binaries by SMBHs is the extremely high velocities of HVSs ejected from galaxy centers [see Eq.~(\ref{eq:v_ej})]. However, HVSs found in the Milky Way have lower velocities. It is unclear whether this discrepancy is due to incomplete HVS search where fast (late-type) HVSs are missed or some unsettled problems for invoking the Hills mechanism as the origin of rpTDEs.
More investigations on the distribution of binary separations and the dependence on the radial distance to the SMBH for different stellar types are needed to resolve the discrepancy. On the other hand, a more complete search of HVSs, especially low-mass HVSs, is necessary for verifying the prediction of the Hills channel.

\begin{acknowledgments}
We gratefully acknowledge the Weihai TDE meeting held in the summer of 2025, which stimulates many valuable discussions. We thank 
Ning Jiang for examining the list of rpTDE candidates, 
Di Wang for pointing out the recent studies on binary separations of stellar binaries in nuclear stellar clusters,  Huan Yang and Xian Chen for discussing the role of dynamical tides in increasing the rpTDE rate, Yang Huang for discussing the prospect of searching for low-mass HVSs with Jiaotong University Spectroscopic Telescope (JUST), and Liang Dai for discussing dynamics of stellar binaries perturbed by a SMBH. We thank the participants of the TDE FORUM (Full-process Orbital to Radiative Unified Modeling) online seminar series for their inspiring discussions. 
\end{acknowledgments}

\appendix

In this Appendix, we summarize in Table.~\ref{table}  the rpTDE candidates and in Table.~\ref{tab:gc_hvs} the confirmed and candidate HVSs in the Milky way reported in the literature.

\begin{deluxetable*}{lccc}
\tablenum{1}
\tablecaption{Properties of rpTDE candidates that have been reported in the literature. \label{table}}
\tablewidth{0pt}
\tablehead{
\textbf{Source}  &  \textbf{Orbital period $T_{\rm obt}$ [yr] } & \textbf{SMBH mass $\log_{10}(M_\bullet/M_\odot)$} & \textbf{Reference} }
\startdata
ASASSN-14ko & $0.3$ & $7.86^{+0.31}_{-0.41}$ & \cite{Payne:2020tfp}\\
eRASSt J045650.3-203750& $0.61$ & $7.0\pm 0.4$ & \cite{Liu:2022avb}\\
AT 2023adr & $1.15$ & $6.95\pm0.23/6.56\pm 0.33$ & \cite{Angus:2026cva} \\ 
AT 2023uqm& $1.2$ & $7.18\pm 0.55$& \cite{Wang:2025oyc} \\
AT 2022dbl & $2.0$  & $6.4\pm0.33$ & \cite{Lin:2024reb} \\ 
AT 2020vdq  & $2.7$  & $5.59\pm0.37$ & \cite{Somalwar:2023sml} \\ 
AT 2021aeuk  &  $3$ & $6.9 \pm ?$ & \cite{Sun:2025yac} \\
AT 2018fyk  & $3.3$ & $7.7\pm0.4$ & \cite{Wevers:2022tsr}\\
AT 2019aalc & $3.8$ & $7.1\pm ?$  & \cite{Veres:2024qcm,Sniegowska:2025esm}\\
AT 2022sxl & $7.2$ & $7.75\pm ?$ & \cite{Ji:2025brt}\\ 
IRAS~F01004-2237 & $10.3$ & $7.4\pm ?$& \cite{Sun2024} \\ 
AT 2019azh & $13.2$ & $6.36\pm 0.43$ & \cite{Yao:2026akx, Wevers:2020hyr}\\
AT 2024pvu & $17.1$ & $7.9\pm 0.4$ & \cite{Yao:2026akx}
\enddata
\tablecomments{Question marks in the third column indicate a lack of properly quantified mass measurement uncertainties. 
In Figs.~\ref{fig:stat} and ~\ref{fig:stat_Hills}, a fiducial uncertainty $\pm 0.6$ dex is used for these events.}
\end{deluxetable*}

\begin{table*}[htbp]
\centering
\tablenum{2}
\caption{Confirmed and candidate hypervelocity stars (HVSs) likely ejected from the Galactic Center (GC).}
\begin{tabular}{lccccc}
\hline
\hline
\textbf{Designation} & \textbf{Spectral Type} & \textbf{Galactocentric Velocity $V_{\rm GC}$ (km s$^{-1}$)} & \textbf{Reference} \\
\hline
SDSS HVSs & B-type & $400-700$    & \cite{Brown:2005ta, Brown2014} \\ 
US 708 & O-type & $\sim 750$ & \cite{Hirsch2005}\\ 
LAMOST-HVS 1 & B-type & $\sim550$ &    \cite{Zheng2014} \\
S5-HVS 1 & A-type & $\sim 1700$ &     \cite{Koposov2020} \\
Marchetti2022 sample & Early-type & 500--700      & \cite{Marchetti2022}\\
\hline
\end{tabular}
\label{tab:gc_hvs}
\end{table*}

\clearpage
\bibliography{sample631}{}
\bibliographystyle{aasjournal}

\end{document}